\documentclass[preprint,amsmath,amssymb,superscriptaddress]{revtex4}

\usepackage{graphicx} 
\usepackage{bm} 

\begin{document}
\newcommand{\ethaffil}{Laboratory of Physical Chemistry and optETH, ETH Zurich, CH-8093 Zurich, Switzerland}
\newcommand{\wuerzburg}{Technische Physik, Universit\"at W\"urzburg, Am Hubland, D-97074 W\"urzburg, Germany}

\title{Near-field imaging and frequency tuning of \\ a high-$Q$ photonic crystal membrane microcavity}

\author{S. Mujumdar}
\altaffiliation[Present address: ]{Tata Institute of Fundamental
Research, Homi Bhabha Road, Mumbai, 400 005, India}
\affiliation{\ethaffil}
\author{A. F. Koenderink}
\altaffiliation[Present address: ]{Center for Nanophotonics, FOM
Institute AMOLF, Kruislaan 407, NL-1098SJ Amsterdam, The
Netherlands} \affiliation{\ethaffil}
\author{T. S\"{u}nner}
\affiliation{\wuerzburg}
\author{B. C. Buchler}
\altaffiliation[Present address: ]{ARC Centre of Excellence for
Quantum Atom Optics, Department of Physics, The Australian National
University, Canberra, Australia}
\affiliation{\ethaffil}
\author{M. Kamp}
\affiliation{\wuerzburg}
\author{A. Forchel}
\affiliation{\wuerzburg}
\author{V. Sandoghdar}
\affiliation{\ethaffil}
\email[]{vahid.sandoghdar@ethz.ch}

\date{\today}

\begin{abstract} We discuss experimental studies of the interaction between a
nanoscopic object and a photonic crystal membrane resonator of
quality factor $Q$=55000. By controlled actuation of a glass fiber
tip in the near field of the photonic crystal, we constructed a
complete spatio-spectral map of the resonator mode and its coupling
with the fiber tip. On the one hand, our findings demonstrate that
scanning probes can profoundly influence the optical characteristics
and the near-field images of photonic devices. On the other hand, we
show that the introduction of a nanoscopic object provides a low
loss method for on-command tuning of a photonic crystal resonator
frequency. Our results are in a very good agreement with the
predictions of a combined numerical/analytical theory.

\end{abstract}

\maketitle

Scanning near-field optical microscopy
(SNOM)~\cite{pohl:84,Lewis:84} is in principle capable of offering
an infinitely high optical resolution. This technique examines the
nonpropagating evanescent optical fields by scanning a subwavelength
probe ~\cite{Sandoghdar-varenna}. In the vast majority of the
applications, it has been assumed that the probe does not perturb
the information obtained from the sample under study. However, a few
reports have shown that the fluorescence properties of dye molecules
could be modified by the SNOM
tip~\cite{Ambrose:94,Gersen:00,Gerhardt:07b}. In this Letter, we
experimentally show and theoretically analyze the influence of a
SNOM tip on the light distribution in a photonic crystal resonator
of high quality factor $Q$. Our results have important implications
for SNOM imaging, controlled tuning of high-$Q$ photonic devices and
optical sensing.

Microcavities have stimulated a great deal of activity in design and
fabrication of new optical systems~\cite{Vahala:03}. For most
applications in quantum optics, integrated optics and optical
sensing it is desirable to increase $Q$ and decrease the mode volume
$V$. As the microcavities become smaller though, the $Q$ often
begins to degrade either because of the relative importance of
surfaces or because of diffraction losses. Resonators based on
photonic crystal (PC) structures offer a good
compromise~\cite{Painter:99,Song:05,Herrmann:06,Tanabe:07}. By
controlling the subwavelength features of PC structures, scientists
have demonstrated their potential for molding the flow of
light~\cite{Soukoulis-edit,joannopoulos-book}. However, the
intrinsic sensitivity of a PC performance to its nanoscopic features
also imposes stringent demands on its fabrication accuracy.
Considering that today's control in nanofabrication is still not
sufficient to produce PCs with the exact design parameters, it is
imperative to 1) characterize the end product and 2) fine tune its
properties after manufacturing. In this work, we address both of
these issues in the near field.

In the past few years, several groups have demonstrated the power of
SNOM and related techniques for imaging light propagation and
confinement in PC
structures~\cite{Sandoghdar-PCbook,Balistreri-00S,Bozhevolnyi-02OC,Kramper:04,Srinivasan:04,Wuest:05,Louvion:06}.
Recently, we theoretically analyzed the influence of a nano-object
on the spectral resonance and the intensity distribution of a PC
microresonator with $Q=13000$. We concluded that depending on the
polarizability of the tip, its influence might no longer be
negligible for cavities of low $V$ and high
$Q$~\cite{Koenderink:05}. We found that in general, the tip could
introduce a frequency shift and a broadening in the resonance of the
PC mode. However, we also showed that the effect of the tip could be
exploited to one's advantage for tuning the microcavity resonance by
a large amount \emph{without} inflicting a significant
broadening~\cite{Koenderink:05}. Indeed, the promise of this
technique has been already recognized and the first experimental
attempts along this line have been pursued by applying silicon tips
to PC cavities with low $Q$s of the order of
500~\cite{Maerki:06,Hopman:06}. However, the observed frequency
shifts were accompanied by substantial broadenings of the cavity
resonances. While these results are of interest for switching
purposes, the large induced loss would suggest that they are not
suitable for tuning high-$Q$ systems.

The design of high-$Q$ PC resonators has witnessed a tremendous
progress in the last five years~\cite{Akahane:03,Srinivasan:03}.
Nevertheless, structures with $Q\gtrsim 10000$ have been available
only from a handful of groups because in addition to a suitable
design, they have to meet stringent fabrication
requirements~\cite{Song:05,Weidner:06,Herrmann:06,Tanabe:07}.
Devices based on silicon have reached extremely high $Q$ values
beyond $10^6$~\cite{Tanabe:07}, but GaAs based structures have been
more difficult to master~\cite{Weidner:06,Herrmann:06}. As depicted
in Fig.~1(a), we have used a line-defect heterostructure cavity
design~\cite{Song:05} realized by connecting crystals with lattice
parameters $a_1 = 410$ nm, $a_2 = 400$ nm and $r/a = 0.23$ in a
221~nm thick GaAs membrane ($\epsilon = 11.39$)~\cite{Skauli:03}. A
collinear W1 waveguide throughout the two lattices resulted in a
mode gap in the transmission band diagram of the structure,
confining light in the longitudinal direction, while the photonic
bandgap confined light in the transverse direction. Two
heterostructures were further created on either side of the cavity
to facilitate incoupling and outcoupling of light. The details of
the fabrication procedure are reported in Ref.~\cite{Herrmann:06}.

\begin{figure}
\includegraphics[width=8.5 cm]{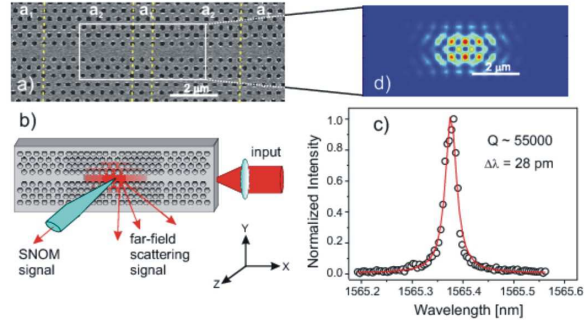}
\caption{\label{fig:simulated} Scanning electron micrograph of the
heterostructure. The yellow lines mark regions with different
lattice constants. $a_1$=410 nm, $a_2$ = 400 nm. (b) Schematics of
the experimental arrangement. (c) The unperturbed cavity resonance
as measured in the far field. The red curve is a Lorentzian fit to
the measured data (black circles). (d) The calculated (FDTD)
intensity distribution on resonance in the region marked by the
white rectangle in Fig.~(a).}
\end{figure}

Linearly polarized laser light from a grating-tunable diode laser
(linewidth~$<300~$KHz, tuning range 1550 nm - 1630 nm) was focussed
onto one end facet of the membrane using a large numerical aperture
lens (NA = 0.68). To facilitate the incoupling, the W1 waveguide of
the cavity was gradually enlarged to a W3 waveguide at the input
facet of the fabricated membrane. As sketched in Fig. 1(b), a small
fraction of light is typically scattered from the surface of a PC
resonator. This light was detected using a lens system and was sent
to a sensitive InGaAs camera or avalanche photodiode (APD) to
perform spectroscopy on the passive resonator. Figure 1(c) displays
an experimental spectrum centered at $\lambda=$1565.38 and a
Lorentzian fit, yielding a linewidth of 28 pm corresponding to
$Q=55000$. We point out that over the months that our measurements
were made, the $Q$ remained the same, but the cavity resonance
frequency shifted, possibly due to environmental changes. As a
result, the exact resonance frequencies discussed in what follows
might differ from one study to the next although they were all
performed on the same PC resonator. Figure 1(d) shows the light
intensity distribution within the cavity on resonance calculated
using the three-dimensional finite difference time domain (FDTD)
method~\cite{Taflove-Hagness,Fan:99}. We employed Liao absorbing
boundary conditions with grid spacings $a/14$ in the lateral and
$a/28$ in the vertical dimensions and used volume averaging of
epsilon to improve the resolution~\cite{Hess:03}.

\begin{figure}
\includegraphics[width=8 cm]{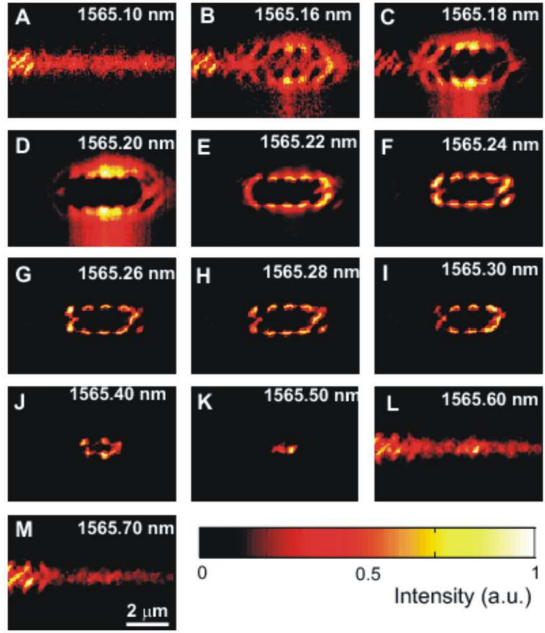}
\caption{\label{fig:exptalscans} Tiles A through M represent SNOM
images of the intensity distribution in the PC structure at
different wavelengths indicated in each image. Each image is
normalized independently according to the color scale shown.}
\end{figure}

As depicted in Fig.~1(b), we employed a SNOM setup to access the
cavity mode in the near field. We used a sharp uncoated heat-pulled
fiber tip (diameter $\sim100$~nm) glued onto a quartz tuning fork.
Shear-force tip-sample distance stabilization was used to maintain
the tip at a distance of about 10 nm from the membrane and to image
the sample topography. The optical fiber tip detected the evanescent
light intensity above the surface using the InGaAs APD at a typical
resolution of 100~nm. Figure 2 shows tiles of recorded images
illustrating the near-field intensity distribution recorded at
various fixed wavelengths close to the cavity resonance. At $\lambda
= 1565.1~$nm (tile A), one can see the light propagating from the
left through the W1 waveguide. At $\lambda = 1565.16~$nm, the light
intensity begins to enter the cavity region, but it also leaks into
the PC structure (tiles B \& C). At $\lambda = 1565.2~$nm (tile D),
most of the intensity is concentrated into a bright elliptical locus
just outside the waveguide. Upon traversing this highest intensity
point, the pattern becomes sharper and gradually collapses into
itself until it ends at a bright point in the center of the cavity
(tiles E-K, $\lambda = 1565.22~$nm$ -1565.5~$nm). Beyond this point
the waveguide mode dominates again (tiles L \& M). Note that the
intensity in each image has been normalized separately so the
waveguide light is overwhelmed in images D-K. These measurements
show the first near-field measurements of light distribution in a
high-$Q$ PC cavity. They also reveal a drastic \emph{spatial}
variation of the intensity distribution within a very narrow
wavelength range. The deviation of the images D-K from that of
Fig.~1(d) hints to the influence of the tip. In other words, at each
wavelength the structure admits light only for a specific locus of
the tip positions.

\begin{figure}
\includegraphics[width=8 cm]{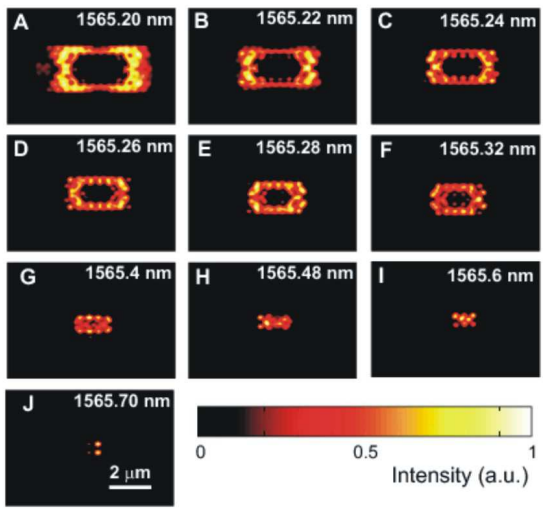}
\caption{\label{fig:simulated} Calculated intensity distribution at
different wavelengths, based on 3D FDTD simulations supplemented
with first-order perturbation calculations to take into account the
impact of the fiber tip. At each wavelength, the structure is
resonant with the incident laser for selected locations of the tip.
The calculation only considered the cavity mode without the
waveguide mode. Each image is normalized independently according to
the color scale shown.}
\end{figure}

In order to verify our experimental findings and their
interpretation, one could perform computationally intensive FDTD
calculations for each tip position. However, a faster and more
instructive check could be achieved by applying the perturbative
treatment described in Ref.~\cite{Koenderink:05}. The frequency
detuning $\Delta \omega$ induced by the tip, or in general a
nanoscopic perturber, is given by,
\begin{eqnarray}
\frac{\Delta\omega}{\omega}&=&
-\frac{\alpha_{eff}}{2}\frac{|E_{0}\bf(r_{\parallel})|^{2}}{\int{\epsilon({\bf
r})|E_{0}|^{2}d{\bf r}}} \exp(-z_p/d)
\end{eqnarray}
where $\alpha_{eff}=3V_{eff}(\epsilon_p-1)/(\epsilon_p + 2)$ is the
perturber's effective polarizability, $z_p$ is its separation from
the sample, and $d$ is the interaction length of the evanescent part
of the cavity mode, found to be 50~nm in FDTD
calculations~\cite{Koenderink:05}. We used the results of the 3D
FDTD calculations for $|E_{0}|^{2}$ shown in Fig.~1(d), the typical
value of $r_p= 50$~nm for an uncoated SNOM tip, and the perturber
effective volume $V_{eff}=\pi r_p^2 d$ to compute the resonance
frequency detuning $\Delta \omega$ at each pixel. Next, we took into
account a Lorentzian profile with a full width at half-maximum of 28
pm (see Fig.~1(c)) for the cavity spectrum and then calculated the
intensity of light at a wavelength of interest. A selection of the
results corresponding to or very close to the wavelengths of the
experimental images in Fig.~2D-M are presented in Fig.~3A-J. We
remark that since the FDTD calculations did not consider the
waveguide mode, the theoretical images in Fig.~3 do not reproduce
Figs.~2A-C or Figs.~2L-M. In fact, Figs.~3I-J indicate that the
experimental images of Figs.~2L-M are dominated by the waveguide
mode. The theoretical and experimental results show a very good
semi-quantitative agreement and confirm that the tip influences the
cavity resonance frequency in a subwavelength position dependent
manner.

\begin{figure}
\includegraphics[width=8 cm]{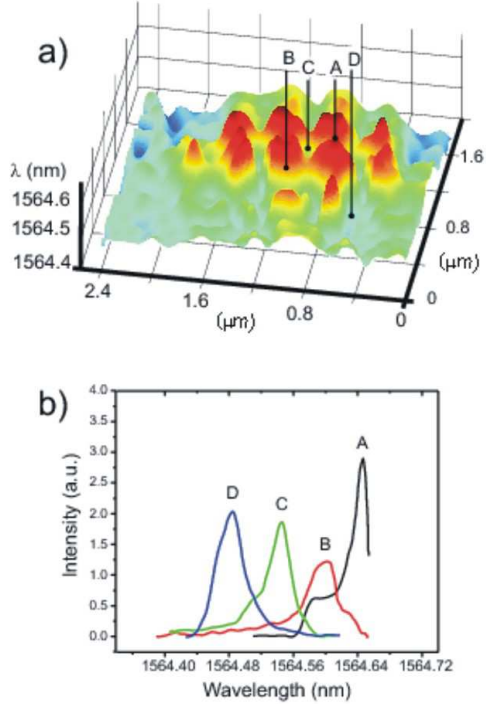}
\caption{\label{fig:pcolor_plus_spectra} a) The peak wavelengths of
the detuned cavity resonance as a function of tip location. b) Four
examples of the local resonance spectra measured through the SNOM
tip at the indicated locations A-D in part (a).}
\end{figure}

Next, we analyzed the complete perturbation landscape by recording
SNOM spectra for tip positions in the x-y plane.  At each tip
position, the laser wavelength was scanned, giving results such as
those shown in Fig. 4(b).  The wavelength of peak intensity for each
tip position is plotted in Fig 4(a) to give a detailed map of the
change in cavity frequency induced by the tip. Over subwavelength
displacements (e.g. see the points `A' to `D'), the resonance shifts
by more than 3 linewidths. As predicted in
Ref.~{\cite{Koenderink:05}, the spatial variation of the tip-induced
resonance frequency shift resembles the distribution of the
intensity in the unperturbed cavity shown in Fig.~1(d). Figure 4(b)
plots examples of spectra recorded at points A-D, revealing that the
resonance lines could deviate from a Lorentzian shape, making it
difficult to assess a quantitative measure of the $Q$ degradation.
However, the data clearly show that the tip-induced broadening is
negligible compared to the frequency shift.

In conclusion, we have used a scanning near-field probe to study the
spatio-spectral features of light in a photonic crystal membrane
cavity of $Q$=55000. The data in Figs. 1, 2, and 3 show that at any
given laser frequency, the resonator admits light only if the probe
is positioned with subwavelength accuracy on a specific locus of
points. We expect similar results also for other cavities with
high-$Q$s and small volumes. The extreme spectral and spatial
sensitivities of such structures to the presence of a subwavelength
object such as a tip means that SNOM images of photonic structures
would have to be interpreted with care. On the other hand, these
sensitivities offer opportunities for a number of applications. In
particular, we have demonstrated that the resonance spectrum of a
microcavity can be manipulated at will by actuating a nanoscopic
object without a notable sacrifice in the resonator $Q$. Such a
nanomechanical actuation could be exploited in compact devices such
as high-$Q$ filters or routers. Furthermore, our findings indicate
that a nanoparticle or a nanoscopic flow of fluid would result in
the spectral modification of the cavity and could be used for
sensing. High-$Q$ cavities are advantageous for these applications
because one can keep the nano-object of interest at fairly large
distances of several tens of nanometers.

We are grateful to C. M. Soukoulis, M. Kafesaki and M. Agio for help
in the early stages of our FDTD work. We are grateful to M. Quack
and G. Seyfang for providing us with the laser source. We
acknowledge the generous support from the Deutsche
Forschungsgemeinschaft (Schwerpunktsprogramm SP1113) and ETH Zurich.

\end{document}